\documentstyle[epsf,sprocl]{article}
\def\Journal#1#2#3#4{{#1} {\bf #2}, #3 (#4)}

\def\PLB{{\em Phys. Lett.}  B}
\def\PRL{\em Phys. Rev. Lett.}
\def\PRD{{\em Phys. Rev.} D}

\def\be{\begin{equation}}
\def\ee{\end{equation}}
\def\bea{\begin{eqnarray}}
\def\eea{\end{eqnarray}}

\begin{document}

\title{PREHEATING AFTER INFLATION}

\author{ Lev Kofman }

\address{Institute for Astronomy,
University of Hawaii\\
2680 Woodlawn Dr., Honolulu, HI 96822, USA}


\maketitle\abstracts{ 
In  inflationary cosmology, the
 particles constituting the Universe are created after inflation
in the process of reheating
 due to the interaction with
the oscillating inflaton field.
We briefly review the basics of the slow 
reheating, and the
stage of fast preheating, when the particles are created explosively in the
regime of parametric resonance. 
The  non-perturbative, out-of-equilibrium  character of
the parametric resonance changes many features of reheating.
 For these  proceedings, 
we will highlight a few   aspects of preheating:
 the structural dependence 
 of the parametric resonance on the inflationary model,
 including $V(\phi)={m^2 \over 2}\phi^2$, ${\lambda\over 4}\phi^4$,
 $1-\cos {\phi \over f}$; ``rescattering'' of  created particles; and
 phase transitions after inflation.}
  
\section{Reheating  after Inflation} 

In modern versions of inflationary cosmology  there is no 
 pre-inflationary hot stage, 
the Universe initially expands quasi-exponentially
in a vacuum-like state without  temperature.
During inflation, all energy is contained
 in the  inflaton field $\phi$ which is 
slowly rolling down to the minimum of its effective potential
$V(\phi)$. When chaotic inflation ends at  $\phi \sim M_p$,
the inflaton   field begins to oscillate 
near the minimum of
 $V(\phi)$ 
with a very large  amplitude  $\phi \simeq {1 \over 10} M_p$.
In this scenario  all the particles constituting the
Universe are created due to the interaction 
with the  oscillating inflaton field.
Gradually, the  energy of inflaton oscillations 
 is transferred into
energy of the  ultra-relativistic particles. Eventually
created particles come to a state of thermal equilibrium
at some temperature $T_r$, which is  called the reheating temperature.
The transitional stage between inflation and the  standard Big Bang
is called the period of reheating after inflation.

To describe  creation of elementary particles from
 the inflaton oscillations,
we shall consider the interaction terms in the 
fundamental Lagrangian.
 The inflaton field $\phi$ may decay
into bosons $\chi$ and fermions $\psi$ due to the interaction  terms $- {
1\over2} g^2 \phi^2 \chi^2$ and
 $- h \bar \psi \psi \phi$, or into its own Bose quanta $\delta \phi$
due to the self-interaction $\lambda \phi^2 \delta \phi^2$,
or due to the gravitational interaction;  
$\lambda$, $ g$ and  $h$ are
small coupling constants. 
In the case of  spontaneous symmetry breaking at a scale $\sigma$, the term
$- {1\over2} g^2 \phi^2 \chi^2$ gives rise to the  
three-legs 
 term  $- g^2 \sigma\phi
\chi^2$.
We will assume for simplicity that the bare masses
of the fields $\chi$ and $\psi$ are very small.
The elementary theory of reheating is based on the perturbative theory
with respect to the coupling constants~\cite{L}.
Inflaton oscillations are interpreted
as a superposition 
 of a number of $\phi$-particles, inflatons,  at rest.
Each inflaton has an energy equal to the frequency of the background
 oscillations. The decay rate of inflatons
(or the particle production rate)
 is given by 
perturbation theory. 
For simplicity, 
  consider  the interaction $-{1 \over 2}g^2 F(\phi) \chi^2$ between
 the {\it classical }inflaton field $\phi$
and the {\it quantum}  Bose field $\chi$
with the eigenfunctions $\chi_{k}(t)\, e^{ -i{{\bf k}}{{\bf x}}}$
 with comoving momenta ${\bf k}$.
 The temporal
part of the eigenfunction   obeys the equation
\begin{equation}
\ddot \chi_k + 3{{\dot a}\over a}\dot \chi_k + {\left(
{{ k^2}\over a^2}   + g^2  F(\phi) \right)} \chi_k = 0 \ ,
\label{2}
\end{equation}
with the vacuum-like initial condition $ \chi_k \simeq {e^{ -ikt} \over
 \sqrt{2k}}$
in the far past.
The WKB  solution of Eq.~(\ref{2}) is
\begin{equation}
a^{3/2}\chi_k(t) \equiv X_k(t) =
{\alpha_k(t)\over \sqrt{2\omega}}\ e^{- i\int^t
\omega dt}
 + {\beta_k(t)\over \sqrt {2\omega}}\ e^{+ i\int^t \omega
dt} \ ,
\label{3}
\end{equation}
where the time-dependent  frequency is $\omega_k^2(t)= {{ k^2}\over a^2}
+ g^2\phi^2$ plus a small correction $\sim H^2, \dot H$; initially
 $\beta_k(t)=0$. For $|\beta_k| \ll 1$, an iterative  solution\\
  $\beta_k \simeq
{\textstyle {1 \over 2}} \int\limits_{ - \infty}^{t} d t'\,{\dot \omega \over
\omega^2}\, \exp{\bigl( -
 2i \int\limits_{ - \infty}^{t'}d t'' \omega( t'')\bigr)} \sim g^2$
gives the standard result of perturbation theory for the
particle occupation number  $n_k= \vert \beta_k \vert^2$.
The integral can be evaluated by
 the  stationary phase  method. For
 the  three-legs interaction $g^2 \sigma \phi \chi^2$,
the perturbative result 
  can be interpreted as   separate  inflatons 
decaying independently of each other into  pairs of $\chi$-particles.
 The resulting
rates of the three-legs
 processes $\phi \to \chi\chi$  and $\phi \to  \psi\psi$
 plus gravitational decay
 are given
by
 \begin{equation}\label{4}
  \Gamma_{ \phi \to \chi \chi} =  { g^4 \sigma^2\over 8
\pi m_{\phi}}\  , \ \
\Gamma_{ \phi \to \psi \psi }  =  { h^2 m_{\phi}\over 8 \pi}\ , \ \
\Gamma_g = {m_\phi^3 \over 8\pi M_p^2}\ .
 \end{equation}
For  the four-legs
 interaction
a    pair of inflatons is
decaying independently  into a pair of $\chi$-particles,
$(\phi \phi \to \chi \chi)$. However, 
the    decay of  massive inflatons in this case 
 is different. The
 transfer of energy from inflatons to the
 created particles 
 rapidly decreases with the expansion of the Universe as
${1\over a^4}{d \over dt}(a^4
\epsilon_{\chi}) \propto a^{-6}.$
 Therefore, the complete
decay of the massive inflaton field  in the theory with
 no spontaneous symmetry breaking or with  no interactions with fermions
 is impossible. 
 
Reheating
completes at the moment $t_r$
 when  the Hubble parameter $H={1 \over 2t}$ becomes equal to $\Gamma_{tot}$.
Assuming  thermodynamic equilibrium 
sets in quickly at the
temperature $T_r$, one can equalize
the energy density of the universe and thermal energy of created
  ultrarelativistic particles at the
 moment   $t_r$. This gives 
the reheating temperature
\begin{equation}
T_r \simeq 0.1 \sqrt{\Gamma M_p}\ ,
\label{5}
\end{equation}
which  is ultimately 
determined by  the total decay rate $\Gamma_{total}$.
In order to get numerical estimates for
$t_r$, $\Gamma_{tot}$ and $T_r$,
 one should know the frequency  of the inflaton oscillations
 and the coupling constants.
 The coupling constants of interaction of
the inflaton field with matter cannot be too large, otherwise the
radiative corrections  alter the shape of the inflaton potential
(unless  SUSY eliminates radiative corrections).
The necessary condition for the decay of inflaton oscillations is
that their frequency is greater than the
effective masses of created particles.
Parameters of the inflaton potential are restricted from the constraints
on amplitude of the cosmological fluctuations.
All together it allows us to put  constraints on the perturbative
reheating.
 The largest  total decay rate   is
$\Gamma < 10^{ -  20} M_p $. It takes at least
$10^{14}$ oscillations  to
transfer the energy of inflaton oscillations into
  the created particles, i.e. this  reheating is very slow.
The   bound on the slow reheating temperature 
is $ T_r <   10^9 \, \mbox{GeV} \ .$
This is a very small temperature, at which neither the standard mechanisms of
baryogenesis in the GUTs  work, nor 
  cosmologically
interesting
heavy  strings, monopoles and textures can be produced.
In the slow reheating scenario with the potential ${m^2 \phi^2 \over 2}$,
the post-inflationary stage of inflaton oscillations with a scalar factor
$a(t) \sim t^{2/3}$
 lasts sufficiently long. It can enhance
the gravitational instability of density fluctuations,
which  leads to the formation of the primordial black holes
for some  specific spectra of the initial density fluctuations.

\section{The Stage of Preheating}

In the elementary theory outlined above, we made significant
oversimplifications, assuming that each inflaton decays independently.
However,  interacting with quantum particles,
inflatons  act not  separately, but as 
  the coherently oscillating 
homogeneous  field $\phi(t)$,
as  follows from  Eq.~(\ref{2}).
The oscillating effective frequency $\omega(t)$
and very large amplitude of the background oscillations $\phi(t)$ can
result in a broad parametric resonance of the ``oscillator'' amplitude
 $\chi_k$.
The smallness of $g^2$ alone does
not necessarily correspond to small occupation number
$\vert \beta_k \vert$, which can grow exponentially.
 Then  the energy is very rapidly
transferred from the inflaton field to other bose fields.
 This process occurs far away from thermal equilibrium, and
therefore it is called  {\it preheating}~\cite{KLS,Shtanov,Boyan95}.
Reheating never completes at the first 
stage of parametric resonance; eventually the resonance becomes narrow and
inefficient. The inflaton field decay is completed and 
 created particles are  settled in thermal equilibrium 
during the subsequent, final  stages of reheating. 
However, the elementary theory 
 should be applied not to the
original inflaton  oscillations, but to the products of its
decay, or to its residual oscillations.
 The short stage of explosive preheating
 in the broad resonance out-of-equilibrium 
regime may have long-lasting effects on
the subsequent evolution of the universe
(compare  to other cosmological epochs where particles
temporarily can be out-of-equilibrium). 
 It may lead to specific nonthermal
phase transitions in the early universe and to the 
 production of 
topological
defects~\cite{KLS96}, 
 it may make possible novel mechanisms of baryogenesis~\cite{KLR,KLS96},
 and it may change the final value of the reheating
temperature $T_r$.

The  non-perturbative character of
the parametric resonance changes many features 
of the  particle
 creation from inflatons. 
 For these  proceedings, 
I will highlight only 
 several interesting aspects of preheating, which we
have learned recently: the structural dependence 
 of the parametric resonance on the theory of inflation $V(\phi)$;
``rescattering'' (mode mixing) effect for created particles; and
phase transitions after inflation. 

\section{Structure of Resonance and Inflaton Potential $V(\phi)$. }

The resonance depends on the form of the background oscillations,
which are defined by $V(\phi)$.
We  consider several popular models of  $V(\phi)$.
 
\subsection{ Resonance in $ {m^2 \over 2}\phi^2$ Inflation}

For the quadratic inflaton potential 
we have   the background inflaton oscillations
$\phi(t) \approx \phi_0(t) \cdot
 \sin{\left( m_{\phi}t \right)}$,
$\phi_0(t)= {M_p \over \sqrt{3\pi}}\cdot{1  \over m_{\phi}t}$.
The scalar factor  is
$ a(t) \approx a_0 t^{2/3}$.
 The amplitude of oscillations is
 decreasing as the universe expands, $\phi_0(t) \sim {0.1 M_p } {a^{- 3/2}}$
Let us consider  ${g^2\over 2}\phi^2\chi^2$ interaction.
The equation for the fluctuations (\ref{2}) with $F(\phi)=\phi^2$
in a new time variable $z=mt$
takes  the form reminiscent of  the  Mathieu equation.
The dimensionless coupling parameter $q= {{g^2 \Phi^2} \over{4 m^2}}$
depends on time via $\Phi^2 \propto a^{-3}$.
Without the expansion of the universe, that equation is exactly
the Mathieu equation.
Since ${m \over M_p} \simeq 10^{-6}$,
we can have  $q \simeq 10^{10} g^2 \gg 1$, so the parametric resonance
is broad.
The regular Mathieu resonance can be described by its
stability/instability chart.
 The Mathieu equation admits
 exponentially  unstable solutions $\chi_k \propto \exp (\mu_k z)$
 within the set of resonance bands, corresponding
to the parametric resonance. For given parameter $q$, there are separated
stability and instability zones.
 For  $q \gg 1$ the parametric resonance is broad,
with the width of the leading  resonance band $ \simeq 0.1 \sqrt{g m \Phi}$ and
with the characteristic exponent $\mu_k$ ranging from zero to
its maximum $0.28$. 
In this case the eigenfunctions
$\chi_k(t)$ have the form of the WKB solution (\ref{2})
 for all $t$ 
except very short time around moments $t_j$, $j=1,2,3, ...$,
 where the background oscillating field
crosses zero, $\phi(t_j)=0$.
\begin{figure}[t]
\centering
 \hskip -0.9 cm
\leavevmode\epsfysize=4.5cm \epsfbox{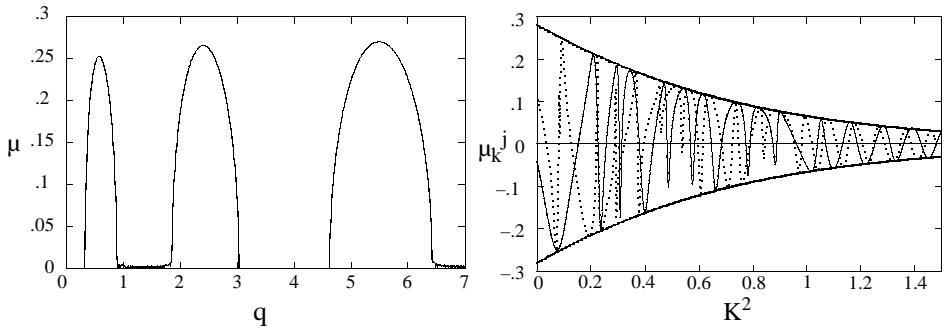 }\\
\

\caption[fig1]{\label{fig:cosmo1}
Left: characteristic exponent $\mu$ of the Mathieu equation as 
function of the parameter $q$ for the fluctuation with  $k=0$.
Right: the characteristic exponent $\mu^j_k$ 
 in an expanding universe
as function of $k$ for $j=5$ and $6$.
 $\mu^j_k$ changes dramatically with $j$ but within the theoretical 
  envelope~\cite{KLS97}.
}
\end{figure}
Non-adiabatic changes of the amplitude  $\chi_k(t)$ occur only
 in the vicinity of $t_j$, which can be described
 with the parabolic scattering.
An essential parameter here is 
the phase $\theta_k =\int^{t_{j+1}}_{t_j }dt \sqrt{k^2+ g^2 \phi^2(t)}$
 accumulating between two successive zeros of $\phi(t)$.
If the parameter $q=const$, the phase $\theta_k$ as function of $k$
is not varying. Then we have a stable 
phase correlation/anticorrelation between successive scatterings
at the parabolic potentials, which produces the sequence of
stability/instability bands, see Fig.~\ref{fig:cosmo1}.

However, in an expanding universe the parameter $q$ is 
 time-depended,  $q={{ g^2 \Phi^2} \over 4 m^2}$:
 $q  \propto j^{-1} $.
For the broad resonance case $q \gg 1$
this parameter can jump over a number of instability bands within a
single oscillation, and the concept of stability/instability bands is
inapplicable here. It is easy to understand that in this case
the parametric resonance is a stochastic process.
Indeed, for   large initial values of $q$,
 the phase rotation
  between $t_j$ and $t_{j+1}$,
$\delta \theta_k  \simeq    {{ \sqrt{q}} \over 8j^2}$
 is much larger than $\pi$ and therefore 
can be considered as a  random number in  the interval $[0, 2\pi]$. 
As a result, there are no separate stability/instability bands,
but each mode can be amplified or deamplified
 with every half   period of  the inflaton oscillation.  
An example of a stochastic $\mu_k$
 is plotted in  Fig.~\ref{fig:cosmo1}.
 Positive and negative occurrences of
$\mu_k$  for $\kappa < \pi^{-1}$   appear in the proportion $3:1$,
and the net effect is  parametric amplification.

\subsection{ Resonance in  ${1 \over 4} \lambda \phi^4$   Inflation}

For the theory with the potential  $V(\phi)={1 \over 4} \lambda \phi^4$
 it is more convenient to express
 the background solutions   via the conformal time
  $\tau = \int {dt\over a(t)}$ and 
conformal field $\varphi=a \phi$ variables:
$\varphi(\tau) \approx   \tilde\varphi  ~
  cn \left( \tau, { 1 \over \sqrt{2}}\right)$,
$ a(\tau)=\sqrt{2 \pi \over 3 }{\tilde\varphi\over M_p}\tau$,
 $\tilde\varphi$ is  the amplitude of oscillations.
The oscillations in this theory  are not sinusoidal, but given by elliptic
 function.
Let us consider a  ${g^2\over 2}\phi^2\chi^2$ interaction of the inflaton.
Eq.~(\ref{2})   for quantum fluctuations $\chi_k$
 can be simplified in this theory. 
Using a conformal transformation of the mode function
$ X_k(t)= a(t)\chi_k(t) $, from 
   Eq.~(\ref{2}) we obtain
\begin{equation}
X_k''  +  {\left(\kappa^2  + {g^2\over \lambda}
   cn^2 \Bigl(\tau,  { 1 \over \sqrt{2}}\Bigr)
 \right)} X_k  = 0 \ ,
\label{10}
\end{equation}
where $\kappa^2={ k^2 \over  \lambda   \tilde\varphi^2}$,
and $q= {g^2\over \lambda}$.
The equation for fluctuations does not depend on the
expansion of the universe and is completely reduced to the similar
problem in Minkowski space-time.
This is a special feature of the
conformal  theory
 ${1 \over 4} \lambda \phi^4  + {1 \over 2} g^2 \phi^2 \chi^2 $.
The mode equation  (\ref{10})
belongs to the class of   Lam\'{e} equations~\cite{KLS,Boyan96,Kaiser}
The combination of parameters  $q=g^2/\lambda$   ultimately
defines the structure of the parametric resonance in this theory.
This means that the condition of  a  broad parametric resonance
does not require a large initial amplitude of the inflaton field,
 as for the quadratic potential.
The strength of the resonance
depends non-monotonically on this parameter. 
The stability/instability chart of the  Lam\'{e} equation (\ref{10})
in the variables
$\left(\kappa^2, {{g^2} \over \lambda} \right)$ was constructed
in \cite{GKLS} (see also P.Greene's contribution to this volume).
On the left panel of Fig.~{\ref{fig:cosmo2} we show
 slices of the Lam\'{e} stability/instability chart as a functions
 $\mu_k$  for several values
of parameter $g^2/\lambda$.

To see how the general theory of the  Lam\'{e} resonance works,
for illustration let us consider  
parametric resonance in the model of
the self-interacting two-component scalar field  $(\phi_1, \phi_2)$
with the effective potential
$V(\phi) = {\lambda\over 4} ( \phi_1^2+ \phi_2^2)^2$.
For the two  component inflaton scalar field,
we always can align the classical background field
with one of the components,
 say  $\phi_1(t)$.
The equations for the quantum fluctuations
for two components $\delta \phi_1$ and $\delta \phi_2$
are different.
The equation for the mode function of fluctuations
in the  direction $\phi_1$ is
\begin{equation}
\ddot {\phi_{1k}}  +    3{{\dot a}\over a}~{\dot \phi_{1k}}+   {\left(
{k^2\over a^2}
   +  3 \lambda \phi_1^2 
 \right)} \phi_{1k}  = 0 \ .
\label{fluc1}
\end{equation}
The equation for the mode function of fluctuations
in the  direction $\phi_2$ is
\begin{equation}
\ddot {\phi_{2k}}  +    3{{\dot a}\over a}~{\dot \phi_{2k}}+   {\left(
{k^2\over a^2}
   +   \lambda \phi_1^2 
 \right)} \phi_{2k}  = 0 \ .
\label{fluc2}
\end{equation}
In some early papers on preheating the iteration series in the small parameter
$\lambda$ was used. This approach is incorrect. Indeed,
using the conformal transformation of the mode functions
$ \varphi_{1k}(t)= a(t)\phi_{1k}(t) $
and $ \varphi_{2k}(t)= a(t)\phi_{2k}(t) $, we can
reduce the equations  (\ref{fluc1}), (\ref{fluc2}) 
to the general equation  (\ref{10}) with particular value of the parameter
$q=3$ and $1$ respectively, so there is no small parameter
associated with the resonance. In both cases $ q =1, 3$ there is
only one instability band, but the strength of the resonance is 
different~\cite{Boyan96,Kaiser,KLS97}.
 The resonance in the ``inflaton'' direction
$\phi_1$ is weak, maximal value of the characteristic exponent 
of the fluctuations $\phi_{1k} \propto e^{\mu_1 \tau}$  is
 $\mu_1 \approx 0.036$; the resonance in the other  direction $\phi_2$
is much   stronger
and broader, $\mu_2\approx 0.147$, see Fig.~\ref{fig:cosmo2}.

\subsection{ Resonance in   Sine-Gordon theory}

\begin{figure}[t]
\centering
 \hskip -0.6 cm
\leavevmode\epsfysize= 4.5cm \epsfbox{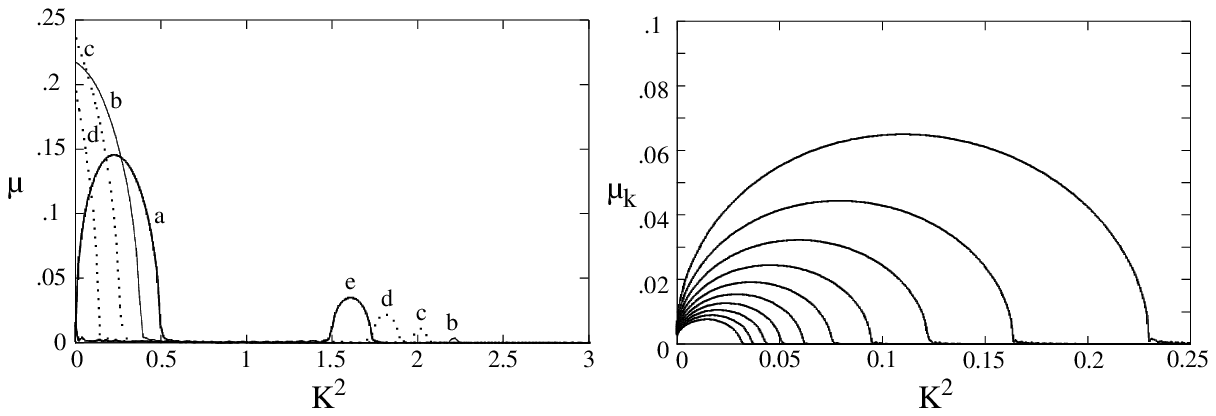}\\
\caption[fig1]{\label{fig:cosmo2}
Left: The characteristic exponent $\mu_k$ for the  Lam\'{e} equation
(\ref{10}) as a function of  $k^2$ (in units of  $\lambda   \tilde\varphi^2$ )
for ${g^2 \over \lambda}=1.0, 1.5, 2.0, 2.5$ and $3.0$, labeled
$a$ through $e$ respectively.
Right:  $\mu_k$ 
as a function of  $k^2$ (in units of $m^2$)
 in the Sine-Gordon theory. Different curves correspond to  different
values of the background misalignment angle, $\theta_0$, from the outer curve
inwards $\theta_0= 1.67, 1.25, 1.0, 0.83, 0.71,$
$ 0.625, 0.56, 0.5,
 0.45, 0.42, 0.38, 0.36.$ }
\end{figure}
The Sine-Gordon potential
$V(\phi) =  m^2  f^2 \left (1 - \cos {\phi\over f  }\right )$ is the prototype
of  the natural inflation scenario, as well as the cosmic axions.
Let us consider small quantum fluctuations $\delta \phi$ around
 oscillating homogeneous background field $\phi(t)$, which can be excited
due to the  self-interaction.
At the moment we neglect the expansion of the universe.
The equation for the temporal part of the eigenfunctions of fluctuations
 $\phi_k(t)$  is 
\begin{equation}\label{sgf}
\ddot {\phi_k}+{\left(k^2 + m_a^2\, \cos {\phi \over f}
\right)} \phi_k  = 0 \ .
\end{equation}
Since the background oscillations are
periodic,  there is parametric amplification of the fluctuations
 $\phi_k(t)$. 
We~\cite{GKS}
 have derived the general analytic  solution  of this 
  equation
for an arbitrary initial amplitude $\phi_0$, or
for an arbitrary  misalignment angle $\theta={\phi_0 \over f}$.
 The characteristic exponent $\mu_k$ as function of 
$k$ is plotted in Fig~(\ref{fig:cosmo2}).
There
 is always only a single resonance band.
The value of  $\mu_k$ strongly depends on the phase
$\theta_0$.
In the important case  of small misalignment 
angle  ${\theta_0} \ll 1$,
the general formula for $\mu_k$
is reduced to the simple expression 
$\mu_k={k \over 4m_a}\sqrt{ {\theta_0^2   } -  { 4k^2 \over m_a^2}}$.
In this case 
 the  maximum  value $\mu_k ={ \theta_0^2\over 16 }$
at  ${k^2 \over m_a^2} = { \theta_0^2\over 8 }$.
The resonance band in this limit is
$0  \leq {k^2 \over m_a^2} \leq { \theta_0^2\over 4 }$.
It is instructive to return to the equation for fluctuations
 (\ref{sgf}), and perform there a small $\theta$
approximation to the $\cos  \theta$ term. 
This gives us the 
Mathieu equation with the parameter
 $q = {\theta_0^2\over 8}$.
We would predict  a set of
resonant bands,  the first resonance band being located  at
${ \theta_0^2\over 8 }  \leq {k^2 \over m_a^2} \leq { 3\theta_0^2\over 8 }$,
 which is shifted from the
actual  instability zone.
The difference between the actual  Sine-Gordon resonance and its 
Mathieu approximation is even more noticeable if we include the
expansion of the universe.
The parametric resonance due to the self-interaction
is not effective both for the natural inflation and for axions,
because the amplitude  
 is quickly $\theta_0 \sim a^{-3/2}$
 decreasing with the
expansion, but not because of the  redshift from the resonance
as  was thought previously.
Certainly, if we 
allow coupling to other bosons, say a $g^2\phi^2\chi^2$-interaction,
 then preheating in  natural
 inflation can be very efficient.

\section{``Rescattering'' of created particles}

So far we have treated the fluctuations $\chi$ of $\delta \phi$
as  test fields in a given background $\phi(t)$ and $a(t)$.
Due to the exponential instability of fluctuations, one expects
their backreaction on the background dynamics is gradually accumulating until
it changes the resonance itself. There are two especially important
effects of backreaction. First, fluctuations may change  the
frequency of background oscillations,
which can be taken into account in the Hartree
 approximation~\cite{Boyan95,KLS}.
The second effect is the production of the inflaton fluctuations,
which occurs due to the interaction of created particles 
 $\chi$ (or $\delta \phi$)  with the inflatons at rest.
One can visualize this process as scattering of 
 $\chi$ particles on the oscillating field  $\phi(t)$,
taking inflatons away from the homogeneous oscillating condensate.
Fluctuations of Bose fields generated with  large occupation number
can be considered as classical waves with gaussian statistics.
Therefore,  all the field $\phi(t)$, $\chi$, $\delta \phi$
can be treated as interacting classical waves.
This opens an avenue to study preheating through lattice numerical
simulations~\cite{KhTk,Prokopec}.

 The theory of ``rescattering''
can be illustrated with the two-component self-interacting  model
$V(\phi) = {\lambda\over 4} ( \phi_1^2+ \phi_2^2)^2$,
where the inflaton field is
identified with the first component $\phi_1$.
In this model we have the cross-interaction term $\lambda \phi_1^2\phi_2^2$.
The rescattering of the $ \phi_2$ particles
on the classical field  $\phi_1$  leads to the production of
 $ \phi_1$ particles in the process $ \phi_{2,\vec k}~\phi_1
\to  \phi_{2,\vec k'}~ \phi_{1, \vec k-\vec k'}$.
The generation of  fluctuations
 $ \phi_1$ due to this process 
can be incorporated in  the mode equation (\ref{fluc2})
with  the scattering term
 $ \lambda \phi_1(\tau)
 \int d^3k' \phi_{2, \vec k-\vec k'}\phi_{2, \vec k'}$.
In conformal variables, 
the fluctuations  $ \varphi_{1k}$ due to the rescattering are given by
the formula (cf.~\cite{KLS97,KhTk}):
\begin{equation}
\varphi_{1,k}(\tau)=-{1 \over {(2\pi)^2\omega_k}}\int_0^{\tau}d\tau'
\sin \omega_k(\tau-\tau')\phi_1(\tau')  d^3k' 
\varphi_{2, \vec k-\vec k'}(\tau')\varphi_{2, \vec k'}(\tau') \ ,
\label{res}
\end{equation}
where $ \omega_k$ is the effective frequency of Eq.~(\ref{10}).
From this it follows that the amplitude $\varphi_{1,k}(\tau)$
grows with time as $e^{2 \mu_2 \tau}$, where $\mu_2=0.147$.
Therefore  ${\langle   \varphi_1^2 \rangle}$
 evolves in time even faster than ${\langle   \varphi_2^2 \rangle}$.
Analysis of Eq.~(\ref{res}) shows that the process 
 $\delta \phi_{2,\vec k}~\phi_1
\to \delta \phi_{2,\vec k'}~\delta \phi_{1, \vec k-\vec k'}$,
strictly speaking,
does not correspond to the ``particle-like'' rescattering, but rather
to the interaction of waves,  mode  mixing.

\section{Phase Transitions from Preheating}

The theory of cosmological phase transitions is one of the main topics in
modern cosmology. Cosmological phase transitions in GUTs, which may happen at 
  $T_c \sim
10^{15}-10^{16}$ GeV, could give rise to primordial monopoles and
other topological defects, which could either kill  popular cosmological
models, or help them.
The possibility of the  phase transitions from preheating 
 is based on the following
idea \cite{KLS96}.
The number of particles created during preheating
 can be very large.
 These particles initially are far away from
thermal equilibrium.
They may change the shape of the effective potential,
which may lead to specific nonthermal phase transitions soon after inflation.
Nonthermal  phase transitions induced by preheating are similar to  the usual
high-temperature phase transitions, but they may occur even in the
 models where the high temperature effects cannot lead to symmetry
restoration. This is because fluctuation amplitudes 
$\langle\phi^2\rangle$ from resonant amplification can be very large.

As an example, let us consider a
two-component scalar field  $(\phi_1,\phi_2)$
with the effective potential
\begin{equation}\label{22}
V(\phi) = {\lambda\over 4} ( \phi_1^2 + \phi_2^2 -{\rm v}^2)^2 \ .
\end{equation}
If we identify the $\phi_1$ component with the ``inflaton'' direction
(as in   section 3.2),
the effective mass of the background field is
$m_{eff}^2=-{\rm v}^2+3\lambda\phi_1^2+3\lambda \langle\delta\phi_1^2\rangle
+ \lambda \langle\delta\phi_2^2\rangle$. 
We shall consider the growth of fluctuations in this model.
At the beginning one can neglect the tachyonic term. Then
we can use the results of
 Sec. 3.2, 
 $\langle\delta\phi_1^2 \rangle     \propto e^{2\mu_1 \tau}$,
 $\langle\delta\phi_2^2 \rangle \propto e^{2\mu_2 \tau}$, $\mu_2 \gg \mu_1$.
However, the mode mixing due to the cross-interaction
leads to the faster growth of fluctuations
 $\langle\delta\phi_1^2\rangle \propto e^{4\mu_2 \tau}$.
If ${\rm v}^2 > 10^{-4} \tilde \varphi^2$, the generation of fluctuations
will be completed before the bifurcation in the evolution of the
 the background field occurs at $\phi_1 \simeq {\rm v}$.
By that moment, we will have the amplitude  of fluctuations 
comparable to that of the background field, 
 $\langle\delta\phi_1^2\rangle \sim \langle\delta\phi_2^2\rangle 
\sim \phi_1^2$.
In this model we have  symmetry restoration alongside with 
 the residua
oscillation of the  background component
$\phi_1(t)$, and what is most important, the formation
of  strings. Notice that the string formation mechanism
is somewhat different from the Kibble mechanism. Lattice simulations of the
self-consistent non-linear  dynamics clearly demonstrate the
formation of topological defects in the model \cite{TKLSK}.

\section*{Acknowledgments}

The reported results are based on the collaborations
 \cite{KLS97,GKLS,GKS,TKLSK}.
This work was supported by NSF Grant No. AST95-29-225.

\end{document}